\documentclass[apl,aip,english,twocolumn]{revtex4} \usepackage[T1]{fontenc}
\usepackage[latin1]{inputenc} \usepackage{babel} \usepackage{txfonts}
\usepackage{epsfig,epsf,psfrag} \usepackage{graphicx} \usepackage{epic,eepic}
\usepackage{color,pstcol} \usepackage{fancyhdr}  \parindent0em
  \def\prl{Phys. Rev. Lett. }

\vspace{-10.0cm} 
\begin{document} 
\title{How to detect a genuine quantum pump effect in
graphene?} \author{Colin Benjamin} \affiliation{ National institute of Science education \&
Research, Bhubaneswar 751005, India }

\begin{abstract} Quantum pumping in graphene has been predicted in recent years. Till date there
have been no experiments indicating a graphene based quantum pump. This is not uncommon as in case
of other non-Dirac behavior showing materials it has not yet been unambiguously experimentally
detected. The reason being that in experiments with such materials the rectification effect
overshadows the pumped current. In this work we answer the question posed in the title by taking recourse to ``strain''. We show that the symmetries of the rectified and pumped currents towards strain reversal can  effectively distinguish between the two.
\end{abstract}

\maketitle

The field of quantum pumping burst into the limelight because of an ingenious experiment performed
in 1999. Applying time dependent magnetic fields to a quantum dot, the authors experimentally showed the
existence of a pumped current in absence of any dc voltage bias\cite{marcus}. However, the pumped
current was seen to be symmetric with respect to magnetic field reversal. This was a rude shock
since pumped currents being dependent on scattering amplitudes unlike two terminal Landauer
conductance which depends on probabilities should possess no symmetry with respect to field
reversal\cite{shutenko,kamenev}. The consensus now is that quantum pumping if at all present is masked by the rectified
currents in the aforesaid experiment\cite{cb0}. Thus in this situation how to isolate a genuine quantum pump effect. The way forward is to check for magnetic field
symmetry of the pumped and rectified currents. The rectified currents are symmetric with respect to
magnetic field reversal while the pumped currents do not possess any definite symmetry with respect
to field reversal. This feature has been exploited in some works\cite{shutenko,kamenev} which dealt with this topic in the
context of non-graphene systems. What about graphene? A lot many works have appeared in the literature about graphene based adiabatic quantum pumping, among the earliest was  Ref.~[\onlinecite{prada}] wherein the importance of evanescent modes was brought about, spin polarized quantum pumping\cite{gra-spin} have been predicted, adiabatic quantum pumping in graphene bilayers\cite{wakker}, in graphene Normal-Superconductor structures\cite{gra-NS} and with magnetic barriers\cite{grichuk}. Finally, noise in graphene quantum pump has been dealt with in Ref.~[\onlinecite{zhu}] and a graphene quantum pump in the non-adiabatic regime has been explored in Ref.~[\onlinecite{non-ad}].  An experimental demonstration of a graphene pump in the non-adiabatic regime has recently been reported in Ref.\cite{connoly}. Since a host of applications of graphene based quantum pumps have been predicted a necessary and fool proof mechanism for its detection should be in place. The purpose of this work is to provide such a tool.

 Applying an external magnetic field to check for symmetries in graphene is unwieldy, in fact this is difficult in any system where magnetic fields aren't a part of the setup abinitio. Further, an external magnetic field may change the magnitude and direction of pumped and rectified currents which has little or no semblance with the original
currents in absence of magnetic fields. Thus because of the twin difficulty (i) external control of
magnetic fields at nanoscale is difficult and fraught with unintended consequences and (ii) an
external magnetic field brought only as a check for pumping/rectification may lead to drastic
changes in the nature of pumping or rectification. We in this letter propose to do some blue sky
thinking and go beyond magnetic fields in graphene. We propose to use strain as a parameter to check on the
nature of pumped/rectified currents.  It has been shown that the Dirac band structure in graphene is fairly insulated upto strain induced elastic deformations of upto $20\%$, beyond which a band gap appears. In this work we are always below the $20\%$ threshold. Thus strain at lower values can be an effective check as it doesnt change the Dirac band structure of graphene. Further strain induced control is easy and flexible since these can be 
managed by gate voltages\cite{ong}. Further more as has been in vogue strain in graphene induces pseudo
magnetic fields\cite{guinea-natphys,pereira,peeters} and via strain lots of new applications for graphene have been predicted\cite{zhai,cao,wu} and strain can genrate a pure bulk valley current too\cite{low}. Therefore strain reversal would automatically imply magnetic field reversal and
therefore as is wont rectified currents would be symmetric with respect to strain reversal while
pumped currents wouldn't. This is the main message of this work.

Now how do these phenomena of rectification or quantum pumping originate. In the experiment which
was supposed to be the first to show quantum pumping what seems to happen is that the time dependent
parameters may through stray capacitances directly link up with the reservoirs. Thus indirectly
inducing a bias which is the origin of rectified currents. Finally, what is quantum pumping? It
refers to an unique way to transport charge without applying any voltage bias. The rectified current
in a two terminal setup is given by\cite{brou_rect}:
 \begin{equation}
I_{rect}=\frac{w}{2\pi}R\int_{S}dX_{1}dX_{2}(C_{1} \frac{\partial G}{\partial X_1}- C_{2}
\frac{\partial G}{\partial X_2}) \end{equation} Herein $R$ is the resistance of circuit path and is
assumed to be much less than the resistance of the mesoscopic scatterer, while $C_1$ and $C_2$ are
stray capacitances which link the gates to the reservoirs, $X_1$ and $X_2$ are the modulated gate
voltages. $G$ is the Landauer conductance which is just the transmission probability (T) of
the mesoscopic scatterer in a two terminal setup. The pumped current into a specific lead in a two terminal system, is in
contrast given as\cite{brou_pump} \begin{eqnarray}
I_{pump}&=&\frac{e}{\pi}\int_{A}dX_{1}dX_{2}\sum_{\beta}\sum_{\alpha \in 1} \Im (\frac{\partial
S_{\alpha \beta}^{*}}{dX_1}\frac{\partial S_{\alpha \beta}^{}}{dX_2})\nonumber\\
&=&I_{0}\sum_{\beta}\sum_{\alpha \in 1} \Im(\frac{\partial S_{\alpha \beta}^{*}}{dX_1}\frac{\partial
S_{\alpha \beta}^{}}{dX_2}) \end{eqnarray} where $I_{0}=\frac{e}{\pi}\int_{A}dX_{1}dX_{2}$. In the
above equation, $S_{\alpha\beta}$ defines the scattering amplitude (reflection/transmission) of the
mesoscopic scatterer, the periodic modulation of the parameters $X_1$ and $X_2$ follows a closed path in
a parameter space and the pumped current depends on the enclosed area $A$ in ($X_1$,$X_2$) 
space. The mesoscopic sample is in equilibrium to start with and for it to transport current one simultaneously varies two system parameters $X_{1}(t)=X_{1}+\Delta X_{1} \sin (wt)$ and
$X_{2}(t)=X_{2}+\Delta X_{2} \sin (wt+\theta)$, herein $\Delta X_{i}$ defines the amplitude of
oscillation of the adiabatically modulated parameter and $\theta$ is the phase difference between the modulated parameters. In the adiabatic quantum pumping regime we
consider the system thus is close to equilibrium\cite{mosk_ref}. 
The main difference between rectified and pumped currents are while the former
are bound to be symmetric with respect to magnetic field reversal (via, Onsager's symmetry) since the
conductance\cite{buttiker} and it's derivatives enter the formula, the pumped currents would have no
definite symmetry with respect to magnetic field reversal\cite{shutenko,kamenev} since they
depend on the complex scattering amplitudes which have no specific dependence on field reversal
unless the scatterer possesses some specific discrete symmetries as shown in Ref.\cite{kamenev}. The
rectified currents in the adiabatic quantum pumping regime considered here differ from that in the
non-linear dc bias regime. In the latter the Onsager symmetry relations are not
obeyed\cite{non-linear} while in the former (from Eq. 1) they are obeyed. Rectification can
also be talked of when a high frequency electromagnetic field is applied to a phase coherent
conductor\cite{falko}. This case also falls into the non-linear regime.

In our analysis we remain in the adiabatic weak pumping regime for both rectified as well as pumped currents.
The weak pumping regime is defined as one wherein the amplitude of modulation of the parameters is
small, i.e., $\delta X_{i} \ll X_{i}$. In this weak pumping regime the rectified currents are given by
\[ I_{rect}=I^{0}_{rect}[C_{1}\frac{\partial T}{\partial X_{1}}-C_{2}\frac{\partial T}{\partial
X_{2}}] \] with $I^{0}_{rect}=we^{2}\sin(\theta)\delta X_{1}\delta X_{2} R/4\pi^{2}\hbar$. $T$ is the
transmittance through the scattering region, which in case of a two terminal set-up is the Landauer
conductance $G$. Similarly the pumped current into say the left lead, $L$ is - \[
I_{pump,\alpha}=I^{0}_{pump}\sum_{\beta}\sum_{\alpha \in L} \Im (\frac{\partial
S^{*}_{\alpha\beta}}{\partial X_{1}}\frac{\partial S^{}_{\alpha\beta}}{\partial X_{2}}) \] with
$I^{0}_{pump}=we\sin(\theta)\delta X_{1} \delta X_{2}/2\pi$, $w$ is the frequency of the applied time
dependent parameter, $\theta$ is the phase difference between parameters and $e$ is the electronic
charge.

In a previous work\cite{cb0} I tried to look beyond these magnetic field symmetry properties and
provide examples wherein the nature or magnitudes of the pumped and rectified currents are exactly
opposite enabling an effective distinction between the two. However in some of the examples used an
artificial condition of exactly equal magnitude of stray capacitances was assumed. Here the
distinguisher is strain applied to a graphene layer and no artificial condition of equal stray
capacitances is required.

\begin{figure}[h] 
 \centerline{ \includegraphics[width=7cm,height=4cm]{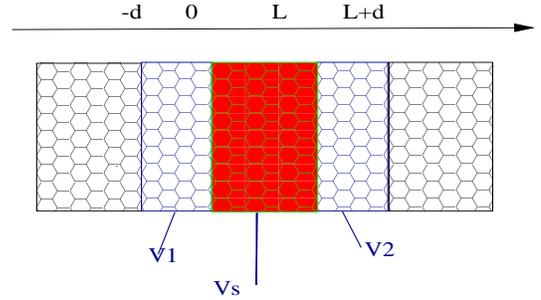}}
\caption{The model system a graphene layer with a strain applied to a specific part. $V_{1} $ and $V_{2}$ are the strengths of the barriers to the left and right of the strained region. In the thin barrier limit they reduce to phases $\chi_{1}$ and $\chi_{2}$. To the left and right of the set-up are shown two contacts $LL$ and $RL$ which are effectively at equilibrium.} \label{scheme}
\end{figure}

 Graphene is a monatomic layer of graphite with a honeycomb lattice
structure~\cite{graphene-rmp} that can be split into two triangular sublattices $A$ and $B$. The
electronic properties of graphene are effectively described by the Dirac
equation~\cite{graphene-sudbo}. The presence of isolated Fermi points, $K_+$ and $K_-$, in its
spectrum, gives rise to two distinctive valleys. In this work we deal with a
normal-insulator-strain-insulator-normal (NISIN) graphene junction. We consider a sheet of graphene
on the $x$-$y$ plane. Strain is induced by depositing graphene onto substrates with regions which
can be controlably strained on demand\cite{ong,neto,ni} by applying a gate voltage. In Fig.~\ref{scheme} we sketch our proposed
system. The strained region is located between $0<x<L$, while the insulators are located on its
left, $-d<x<0$, and on its right, $L<x<L+d$. The normal graphene planes are to the left-end, $x<-d$,
and to the right-end, $x>L+d$.

For a quantitative analysis we describe our system by the Dirac equation in presence of strain that
assumes the form~\cite{neto} \begin{equation} H=v_{F} \Psi^{\dagger} \left[\begin{array}{cc}
{\mathbf \sigma}\cdot({\mathbf p}-\frac{1}{v_{F}}{\mathcal A})& 0\\ 0& -{\mathbf
\sigma}\cdot({\mathbf p}-\frac{1}{v_{F}}{\mathcal A})\end{array}\right]\Psi,
\label{eq:H}\end{equation} where $E$ is the excitation energy, $\Psi$ is the wavefunction.
 
Here $\hbar, v_{F}$ (set equal to unity hence forth) are the Planck's constant and the energy
independent Fermi velocity for graphene, while the $\sigma$'s denote Pauli matrices that operate on
the sublattices $A$ or $B$. Eq.~\ref{eq:H} is valid near the valleys $K$ and $K'$ in the Brillouin
zone and $\Psi=[\psi_{K}^{A}({\mathbf r}), \psi_{K}^{B}({\mathbf r}),\psi_{K'}^{B}({\mathbf
r}),\psi_{K'}^{A}({\mathbf r})]^{\dagger}$ is a spinor containing the electron fields in each
sublattice and valley. In a specific valley the Hamiltonian is $H=v_{F}{\mathbf \sigma}\cdot
({\mathbf p}-\frac{1}{v_{F}}{\mathcal A})$. The electron dynamics in a specific valley is then
determined by the Dirac equation in presence of a gauge field $\mathcal A$. In Ref.[\onlinecite{neto}], using a tight binding model a theory for uniaxial strain in graphene has been developed.

The gauge field $\mathcal A$ is because of the strain induced modulation of the nearest neighbor
hopping parameter $t$. We write $t=t+\delta$ where $\delta$ is the strain induced modulation. The
complex space dependent vector potential ${\mathcal A}={\mathcal A}_{x}-i{\mathcal A}_{y}$ is given by-
${\mathcal A}=\sum_{n}\delta e^{i \mathbf{K \cdot n}} $. Here $n$ is the lattice index in tight binding
model. The strain induced modulation as shown in Fig. 1 happens in the region $0<x<L$ and the
horizontal hopping is modulated by an amount $\delta$ over a finite region of length $L$. Thus
vector potential $\mathcal{A}=\delta \Theta(x)\Theta(L-x)$ is along y direction which coincides with
direction of translational invariance, $\Theta$ being the Heaviside function.
 
Let us consider an incident electron from the normal side of the junction ($x<-d$) with energy $E$.
For a right moving electron with an incident angle $\phi$ the eigenvector and corresponding
momentum reads \begin{equation} \psi_{+}=[1,e^{i\phi}]^{T}e^{i p\cos\phi x},\,\,\, p=(E+E_{F}).
\end{equation} A left moving electron is described by the substitution $\phi \rightarrow \pi -\phi$.

Since translational invariance in the $y$-direction holds the corresponding component of momentum is
conserved. In the insulators, $-d<x<0$ and $L<x<L+d$, the eigenvector and momentum of a right moving
electron are given by \begin{equation} \psi_{iI+}=[1,e^{i\phi_{i0}}]^{T}e^{i p_{iI}\cos \phi^{}_{i0}
x}, p^{}_{iI}=(E+E_{F}-V_{i}), \end{equation} with $i=1,2$ .
 The
trajectory of the quasi-particles in the insulating region are defined by the angles $\phi_{i0}$.
These angles are related to the injection angles by \begin{eqnarray} \sin \phi_{i0}/\sin
\phi=(E+E_{F})/(E+E_{F}-V_{i}) \label{eq-theta} \end{eqnarray} Here, we adopt the thin barrier limit\cite{sengupt-gra}
defined as, $\phi_{i0}, \mbox{ and } d \rightarrow 0,$ while $V_{i}\rightarrow\infty$, such that
$p_{iI}d \rightarrow \chi_{i}$.
In the strained graphene layer, ($0<x<L$), the possible wavefunctions for transmission of a
right-moving electron with excitation energy $E>0$ read \begin{eqnarray}
\Psi_{S+}&=&[1,e^{i\xi}]^{T} e^{iqx}, \end{eqnarray} with $q=(E+E_{F})\cos(\phi)+\delta$ and $\xi=\sin^{-1}(((E+E_{F})(\sin(\phi))-\delta)/(E+E_{F}))$. To solve the scattering problem, we match the wavefunctions at four
interfaces: $\psi|_{x=-d}=\psi_{1I}|_{x=-d},$ $\psi_{1I}|_{x=0}=\Psi_{S}|_{x=0},$
$\Psi_{S}|_{x=L}=\psi_{2I}|_{x=L},$ and $\psi_{2I}|_{x=L+d}=\psi|_{x=L+d},$ where, starting with
normal graphene at left, $\psi=\psi_{+}+s_{11}\psi_{-},$
$\psi_{iI}=p_{i}\psi_{iI+}+q_{i}\psi_{iI-}+m_{i}\psi_{iI+}+n_{i}\psi_{iI-}, i=1,2,$
$\Psi_{S}=p_{S}\Psi_{S+}+q_{S}\Psi_{S-}+m_{S}\Psi_{S+}+n_{S}\Psi_{S-},$ and finally for normal
graphene at the right, $\psi=s_{12}\psi_{+}$. Solving these equations leads to the amplitude of
reflection $s_{11}$, amplitude of electron transmittance$s_{12}$.

After this we are in position to calculate the pumped and rectified currents as follows:

In our case $X_{1}$ is the strength of first thin barrier $\chi_1$ and $X_{2}$ that for second thin
barrier $\chi_{2}$. To invoke pumping in our proposed system we modulate the strengths
($\chi_{1}=\chi_{10}+\chi_{p}\sin(wt)$) and ($\chi_{2}=\chi_{20}+\chi_{p}sin(wt+\theta)$). Herein $w$
is the pumping frequency and $\theta$ is the phase difference between the two modulated parameters. Thus in this adiabatic pumping regime the system is close to equilibrium.


The first issue we tackle is the conductance. We integrate the transmission probability over all
angles of incidence. \begin{equation} G=\int_{-\frac{\pi}{2}}^{\frac{\pi}{2}}\!\!\! d\phi \cos\phi |s_{12}|^{2}. \end{equation} 

\begin{figure} 
\vskip 0.4in
\centerline{
\includegraphics[width=8cm,height=5cm]{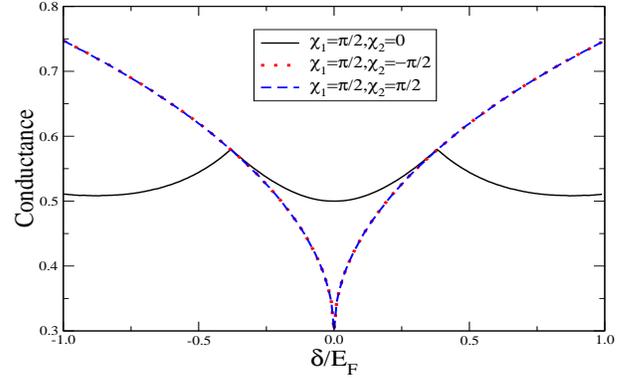}} \caption{Strain reversal symmetry of the
conductance. Parameters are $E_{F}=1.0, E_{F}L=2.0$.}  \label{fig:G-strain} 
\end{figure}

\begin{figure}
 \vskip 0.4in 
\centerline{
\includegraphics[width=8cm,height=5cm]{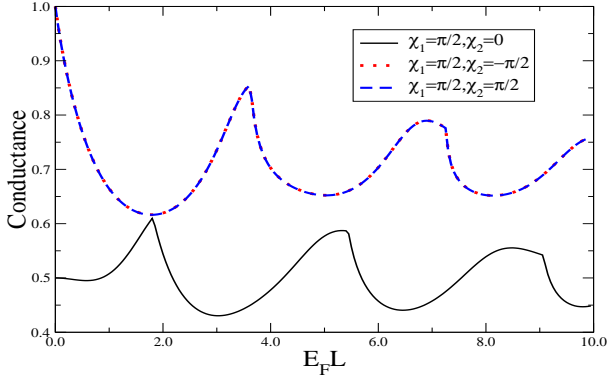}} \caption{Conductance as function of dimensionless length($E_{F}L$) of strained layer. Parameters are $E_{F}=1.0, \delta=0.5$.} \label{fig:G-L} 
\end{figure}
\begin{figure}[h]
\vskip 0.4in \centerline{
\includegraphics[width=8cm,height=5cm]{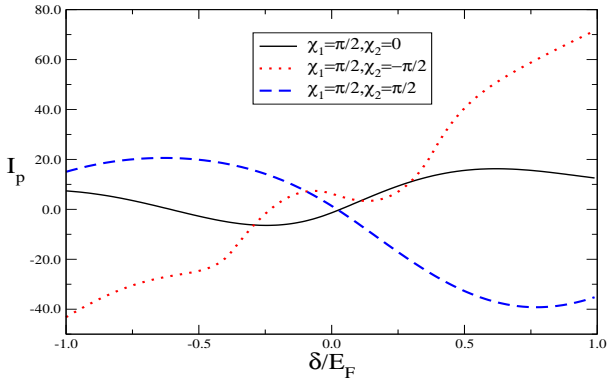}}
 \caption{Nonsymmetry of the pumped
currents $I_{p}$ to strain reversal. Parameters are $E_{F}=1.0, E_{F}L=2.0$.} \label{fig:pump-strain}
 \end{figure}

\begin{figure}[h]
\vskip 0.4in 
\centerline{\includegraphics[width=8cm,height=5cm]{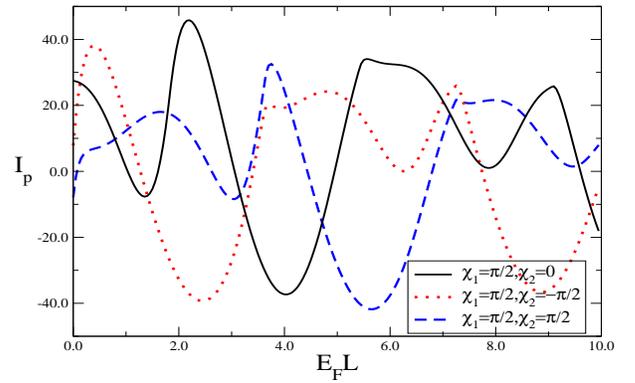}} 
\caption{Pumped currents as function of dimensionless length ($E_{F}L$) of strained layer. Parameters are $E_{F}=1.0, \delta=0.5$.} 
\label{fig:pump-L}
\end{figure} 
The conductance is symmetric under strain reversal as can be seen from Fig.\ref{fig:G-strain}. We take electron excitation energy $E=0$ throughout and $E_F$ is as mentioned in the figure. That
implies rectified currents are symmetric too. It is also periodic as function of the strength of the
insulating barrier's $\chi_{i}$'s (not plotted here). We also plot the conductance
Fig.~\ref{fig:G-L} as function of the length of strained region, the conductance is damped for
increasing lengths. However, one must point out periodic revivals in the conductance because of a
Fabry-Perot type interference which occurs due to the double barrier set up we have here. Next for
the pumped currents, we adiabatically modulate the strengths of the two thin barriers on either side
of the strained region, $\chi_{1}, \chi_{2}$. In the weak pumping regime we have- 
\begin{equation}
I_{p}=\int_{-\frac{\pi}{2}}^{\frac{\pi}{2}}\!\!\! d\phi \cos\phi I_{pump}. 
\end{equation}

with $I_{pump}=I_{0} \sum_{\beta}\sum_{\alpha \in LL}\Im(\frac{\partial S_{\alpha
\beta}^{*}}{d\chi_1}\frac{\partial S_{\alpha \beta}^{}}{d\chi_2})$. In the plots below we plot the
normalized pumped currents $I_{p}$, i.e. divided by $I_0$. We see unlike the conductance the pumped
currents Fig.~\ref{fig:pump-strain}, as is wont are non-symmetric with respect to strain reversal. Since
they are uniquely determined by amplitudes and not probabilities. We also plot the pumped
currents Fig.~\ref{fig:pump-L} as function of the length $L$ of the strained region and not unlike the
conductance, pumped currents are damped too. Ofcourse they are not continuously damped these are interspread with periodic revivals implying
role of Fabry Perot interference because of the two barriers on either side of the strained layer.
A comparison of Figs.~\ref{fig:G-L} and \ref{fig:pump-L} shows another distinction between pumped and rectified currents. One can clearly see the conductance is exactly equal for $\chi_{1}=\chi_{2}=\pi/2$ and  $\chi_{1}=-\chi_{2}=\pi/2$ but the pumped currents are completely different as function of strain length.

The experimental
realization of this structure is not at all difficult, since double barrier structures in graphene
have been experimentally realized, see the recently published work\cite{novo-natcomm} for more details. The only other thing necessary is to have two ac dependent gate voltages to modulate the strength of the double barrier structure. After this we modulate the strain
applied by application of a gate voltage again. If this condition is realized then this very simple
structure will be a very good identifier of a genuine quantum pump effect if present.

\end{document}